# CAN STANDARD COSMOLOGICAL MODELS EXPLAIN THE OBSERVED ABELL CLUSTER BULK FLOW?

MICHAEL A. STRAUSS

School of Natural Sciences, Institute for Advanced Study, Princeton, NJ 08540

RENYUE CEN AND JEREMIAH P. OSTRIKER

Princeton University Observatory, Princeton, NJ 08544

TOD R. LAUER

Kitt Peak National Observatory, National Optical Astronomy Observatories,[1]
P. O. Box 26732, Tucson, AZ 85726

AND

MARC POSTMAN

Space Telescope Science Institute,[2] 3700 San Martin Drive, Baltimore, MD 21218

**ABSTRACT**

Lauer & Postman (LP) observe that all Abell clusters with redshifts less than 15,000 km s$^{-1}$ appear to be participating in a bulk flow of 689 km s$^{-1}$ with respect to the Cosmic Microwave Background. We find this result difficult to reconcile with all popular models for large-scale structure formation that assume Gaussian initial conditions. This conclusion is based on Monte-Carlo realizations of the LP data, drawn from large Particle-Mesh $N$-body simulations for six different models of the initial power spectrum (Standard, Tilted, and $\Omega_0 = 0.3$ Cold Dark Matter, Hot Dark Matter, and two variants of the Primordial Baryon Isocurvature model). We have taken special care to treat properly the longest-wavelength components of the power spectra. The simulations are sampled, "observed," and analyzed as identically as possible to the LP cluster sample. Bulk flows with amplitude as large as that reported by LP are not uncommon in the Monte-Carlo datasets; the distribution of measured bulk flows before error bias subtraction is roughly Maxwellian, with a peak around 400 km s$^{-1}$. However, the $\chi^2$ of the observed bulk flow, taking into account the anisotropy of the error ellipsoid, is much more difficult to match in the simulations. The models examined are ruled out at confidence levels between 94% and 98%. The $1\,\sigma$ error in the amplitude of the LP bulk flow is $\sim 240$ km s$^{-1}$, thus any model that has *intrinsic* flows of less than 480 km s$^{-1}$ on the scales probed by LP scales can be ruled out at a similar level. The LP flow of $\sim 700$ km s$^{-1}$ can thus only be explained as a chance $2 - 3\,\sigma$ event under existing theories.

## 1. INTRODUCTION

With the development of accurate distance indicators for galaxies in the last fifteen years, it has become possible to measure the peculiar velocities of galaxies superposed on the Hubble flow (cf, Burstein 1990 for a review). As the data set of peculiar velocities has grown and the distance indicators have become more refined, it has become increasingly apparent that the peculiar velocity field is a powerful tool for cosmological applications. We may use the observed velocity field to constrain the primordial power spectrum: under the hypothesis of gravitational instability, the peculiar velocity field on large scales is directly related to the power spectrum on these scales, and is largely free of the uncertainties of the relative distribution of galaxies and mass. Indeed, in linear perturbation theory, the root-mean-square bulk flow of a sphere of radius $R$ is given by (Peebles 1993):

$$\left\langle V^2 \right\rangle^{1/2} = \frac{H_0 f(\Omega_0, \Lambda_0)}{2^{1/2}\pi} \left( \int dk\, P(k) \widetilde{W}^2(kR) \right)^{1/2} \quad , \qquad (1)$$

where

$$\widetilde{W}(x) \equiv \frac{3 j_1(x)}{x} \quad , \qquad (2)$$

$$f(\Omega_0, \Lambda_0) = (\Omega_0^{0.6} + \Lambda_0/30) \quad , \qquad (3)$$

$j_1$ is the first-order spherical Bessel function, and $P(k)$ is the mass power spectrum. The quantity $\Lambda_0$ represents the contribution of a Cosmological Constant to curvature; the approximation in Equation 3 is from Martel (1991). Compare this expression with the corresponding expression for the rms mass fluctuations within spheres of the same radius:

$$\left\langle \left(\frac{\delta M}{M}\right)^2 \right\rangle^{1/2} = \frac{1}{2^{1/2}\pi} \left( \int k^2 dk P(k) \widetilde{W}^2(kR) \right)^{1/2} \quad ; \tag{4}$$

for a given volume, the bulk flow is sensitive to the power spectrum at smaller values of $k$, and therefore larger spatial scales, than are the mass fluctuations.

Measurements of the bulk flow of galaxies have been done on ever-larger scales over the years (Burstein 1990; Dekel 1994). Dressler et al. (1987) reported a bulk flow of 600 km s$^{-1}$ on 60 $h^{-1}$ Mpc scales from a peculiar velocity survey of elliptical galaxies, apparently ruling out a wide class of models (e.g., Vittorio, Juszkiewicz, & Davis 1986; Górski et al. 1989). However, the effective volume probed by the observations was smaller than originally claimed (Kaiser 1988), lessening the conflict with models. Recent direct comparisons of observed bulk flows with models include Mucciacia et al. (1993) and Tormen et al. (1993). Dekel, Bertschinger, & Faber (1990) developed a method to smooth the peculiar velocity data, reducing the noise and allowing the measurement of a true volume-averaged bulk flow; Bertschinger et al. (1990) found a bulk flow of $388 \pm 67$ km s$^{-1}$ within a sphere centered on us of radius 4000 km s$^{-1}$. Recently, Courteau et al. (1993) (see also Dekel 1994) combined analyses of several large samples of peculiar velocity data, measuring a volume-averaged bulk flow of $360 \pm 40$ km s$^{-1}$ on a scale of 6000 km s$^{-1}$.

Measurements of bulk flows on large scales can also be viewed as a failure to detect the reflex motion of the Local Group. In the rest frame of the Local Group barycenter, the Cosmic Microwave Background (CMB) shows a dipole moment implying a motion of 622 km s$^{-1}$ towards $l = 277°, b = 30°$ (Smoot et al. 1991). The velocity field of which the Local Group motion is a part is expected to have a finite coherence length, thus if one looks at peculiar velocities of galaxies relative to the Local Group on scales larger than this, one expects merely to see the negative of the Local Group motion. Although Aaronson et al. (1986) claimed to have seen this reflex from observations of ten clusters of galaxies, bulk flow measurements using more complete samples have not yet seen the convergence of the velocity field.

In this paper, we will explore the consequences for cosmological models of the recent results of Lauer & Postman (1994, hereafter LP), who searched for this reflex motion in their peculiar velocity survey of 119 Abell (1958) and Abell, Corwin, & Olowin (1989) clusters. The LP sample is volume-limited and includes all clusters with measured heliocentric redshifts less than $z = 0.05$, or 15,000 km s$^{-1}$, or 150 $h^{-1}$ Mpc. LP used the luminosity of Brightest Cluster Galaxies (BCG) as a distance indicator, a technique first developed by Sandage (1972) and Gunn & Oke (1975). Hoessel (1980) refined the BCG distance indicator, finding a relation between the total luminosity of a BCG and the logarithmic slope $\alpha$ of the integrated luminosity profile. LP and Postman & Lauer (1994) reinvestigated the Hoessel relationship, finding a quadratic relation between $\alpha$ and $L$, the $R_c$ band luminosity within a 10 $h^{-1}$ kpc radius circular aperture. The residual scatter is 0.24 magnitudes, corresponding to an error in distance of each cluster of $0.24/(2 - \alpha) = 0.17$ mag for the mean value of $\alpha = 0.57$. They fit their data simultaneously for the parameters of the $L - \alpha$ relation and a bulk flow. They did not find the expected reflex motion indicative of convergence of the velocity field, but rather measured a bulk flow of 806 km s$^{-1}$ in the CMB rest frame, which, when corrected for error biasing, becomes $689 \pm 178$ km s$^{-1}$, differing from zero at the $4\sigma$ confidence level.

Naïve calculations via Equation 1 show that this result rules out all currently popular cosmological scenarios based on gravitational instability and Gaussian initial conditions at a high significance level. Indeed, in linear theory, the distribution of bulk flows on a given scale is Maxwellian, with standard deviation given by Equation 1. Thus in the Standard CDM model (see below) for which Equation 1 predicts a bulk flow of only 123 km s$^{-1}$ (see Table 2 below), there is much less than one 15,000 km s$^{-1}$ radius volume in the observable universe with a bulk flow as large as observed.

However, such a calculation completely ignores measurement errors. In this paper, we compare these data with models directly, using Monte-Carlo simulations of the LP data set drawn from $N$-body models of a variety of cosmological scenarios. Our motivations for carrying out this more detailed treatment are several-fold:

1. The distance indicator that LP used has appreciable scatter. The average scatter is 17% in distance, but the error on any given BCG varies, depending on $\alpha$; the effect of this scatter can be quantified only approximately using analytic techniques.

2. The LP velocity field is measured for clusters of galaxies, and little is known either observationally or theoretically about the velocity field of clusters relative to field galaxies on large scales (cf, Gramann, Cen, & Bahcall 1994). In particular, clusters in $N$-body simulations, and probably in the real world, do not form at random positions but rather at the highly special places where three caustics intersect (e.g., Cen & Ostriker 1993c). This argues for use of a fully nonlinear simulation in which clusters are identified individually.

3. Equation 1 assumes a strict volume-weighting of the velocity field over a sphere. Although the LP sample is volume-limited, it exhibits strong clustering on small scales (reflecting the cluster-cluster correlation function) and is strongly affected by the zone of avoidance at low Galactic latitudes. Moreover, the anisotropy of the sample causes the error in the detected bulk flow in a given direction to be a strong function of that direction.

4. LP use the sample to solve for the best-fit bulk flow and the distance-indicator relation simultaneously using an iterative technique. This introduces a small geometric bias which they then calibrate and remove in a statistical way. More important, it also introduces a covariance between the bulk flow and the parameters of the $L - \alpha$ relation which is difficult to quantify analytically but can be allowed for in a Monte Carlo simulation.

5. All the above effects may introduce a non-Maxwellian tail into the distribution of the measured bulk flow. It is indeed the tail that we are interested in looking at, and we can quantify it properly only with the use of Monte-Carlo simulations.

The outline of this paper is as follows: In §2, we describe the $N$-body models that were used in this analysis, together with a discussion of the velocity field of clusters relative to the dark matter particles in the $N$-body models. §3 describes our Monte-Carlo simulation technique, and §4 presents results. Our conclusions are presented in §5. Initial results of this investigation are presented in Strauss, Cen, & Ostriker (1994).

## 2. $N$-BODY SIMULATIONS

### 2.1 The Power Spectra

We assume throughout this paper that structure was formed by gravitational instability from initially small-amplitude perturbations with random phases. In this paradigm, only the power spectrum of the initial density field, and the present amplitude of the fluctuations, need be specified in order to determine the statistical properties of the velocity field, as Equation 1 implies. Although much progress has been made in recent years to narrow the range of cosmological models that will fit existing data, there is still a great deal of freedom. With this in mind, we have carried out simulations using six different initial power spectra which cover the full range of currently popular models. All models (with the exception of HDM) are normalized to the observations of CMB anisotropy at $10°$ scales by the Cosmic Background Explorer (COBE; Smoot et al. 1992), interpreted as pure potential fluctuations. The power spectra are shown in Figure 1, with parameters summarized in Table 1, and are as follows:

1. Standard Cold Dark Matter (SCDM), with $\Omega_0 = 1$, Hubble constant $H_0 = 50$ km s$^{-1}$ Mpc$^{-1}$, and fractional rms density fluctuations within $8\,h^{-1}$ Mpc radius spheres of $\sigma_8 = 1.05$ ($\sigma_8$ is just the inverse of the bias factor, as it is often defined). This model assumes a primordial spectral index $n = 1$. The transfer function was taken from Bardeen et al. (1986).

2. Hot Dark Matter (HDM), with $\Omega_0 = 1$, $H_0 = 50$ km s$^{-1}$ Mpc$^{-1}$, and $\sigma_8 = 0.86$, is a model with a much larger ratio of large to small-scale power than standard CDM. If this model is normalized to the COBE fluctuations, galaxy formation in this model happens very late (Cen & Ostriker 1993a), and so we use a normalization that is $1\sigma$ above the COBE normalization.

3. Primordial Baryon Isocurvature (PBI), with $\Omega_0 = 0.2$, ionization fraction $x = 0.1$, isocurvature spectral index $m = -0.5$, $H_0 = 80$ km s$^{-1}$ Mpc$^{-1}$, and $\sigma_8 = 0.9$. These parameters have been chosen to match a variety of observations on various scales (Cen, Ostriker, & Peebles 1993).

4. $\Omega_0 = 0.3$ CDM (LCDM), with $\Omega_0 = 0.3$, $\Lambda_0 = 0.7$, $H_0 = 67$ km s$^{-1}$ Mpc$^{-1}$, $n = 1$, and $\sigma_8 = 0.67$. This model, also normalized to COBE, has a power spectrum with a peak on larger scales than does SCDM, and thus has relatively more large-scale power. It has been invoked to explain observations of the large-scale distribution of galaxies (Baugh & Efstathiou 1993; Peacock & Dodds 1994) and the distribution and mass function of clusters (Bahcall & Cen 1992; Cen, Gnedin, & Ostriker 1993), although because the amplitude of bulk flows scales roughly as $\Omega_0^{0.6}$ (Eq. 3), it does not give larger bulk flows on large scales. Its power spectrum is quite similar to that of TCDM. The value of $\Lambda_0$ adopted is near the upper limit of that permitted by gravitational lensing constraints (cf, Fukugita & Turner 1991; Maoz & Rix 1993).

5. Tilted CDM (TCDM), which assumes a primordial power spectrum with $n = 0.7$. The normalization is again set by COBE, at $\sigma_8 = 0.5$; otherwise, parameters are as in SCDM above. This model was suggested by Cen *et al.* (1992) as a variant of SCDM (see also Lidsey & Coles 1992; Lucchin, Matarrese & Mollerach 1992; Liddle, Lyth & Sutherland 1992; Adams *et al.* 1993; Cen & Ostriker 1993b). The TCDM model decreases the amount of power on small scales relative to SCDM, and increases it on very large scales.

6. Primordial Baryon Isocurvature (New PBI) with $\Omega_0 = 0.3$, $\Lambda_0 = 0.7$, ionization fraction $x = 0.1$, $m = -1.0$, $H_0 = 50$ km s$^{-1}$ Mpc$^{-1}$, and $\sigma_8 = 1.02$. This model fits a suite of observational constraints, including the cluster luminosity function and gas to mass ratio (Cen *et al.* 1994) and the cluster correlation and mass functions (Cen & Bahcall 1994). The properties of the model on smaller scales have not yet been studied. It may in fact be ruled out already by measurements of CMB fluctuations on scales of $1°$, but the observational situation on these scales is controversial. On still smaller angular scales, the amplitude of the CMB fluctuations are dependent on the thermal history of the model; they will be investigated elsewhere (Cen *et al.* 1994).

All models assume Gaussian initial conditions, and all except the two PBI models (which assume nearly isothermal fluctuations) assume adiabatic fluctuations. A Particle-Mesh code was used to simulate each model, using $1.56 \times 10^7$ particles within a box of $800h^{-1}$ Mpc on a side (see Cen 1992 for details). A single simulation was generated for each power spectrum above. From each box, the 16,000 most massive clusters in the volume are identified as peaks in the density field following Bahcall & Cen (1993), and Local Group candidates are identified following Strauss *et al.* (1993, hereafter SCO). The resulting cluster volume density is roughly twice that of Abell (1958) Richness Class 0 clusters.

Perhaps the only model currently popular which we do *not* simulate is that of Mixed Dark Matter (Klypin *et al.* 1993 and references therein). It is intermediate between HDM and SCDM on cluster scales. However, we saw in Figure 4 that the bulk flows on any scale are sensitive to the power spectrum on larger scales, where the COBE normalization guarantees that the Mixed Dark Matter is indistinguishable from SCDM. Thus we expect the results using this model to be similar to those for SCDM.

## 2.2. *The Cluster Velocity Field*

Before making realizations of the LP dataset itself, we can use the $N$-body simulations to answer questions about the velocity field of clusters compared to that of the field, building on work of Gramann *et al.* (1994). In the SCDM simulation, we compared the peculiar velocity of each cluster with that interpolated via cloud-in-cell (CIC) from the velocity field of a Cartesian set of grid points separated by $20\,h^{-1}$ Mpc. The velocity field on the grid was calculated as the ratio of the momentum and mass fields as defined by CIC. The results are shown in Figure 2a, for the X-component of the peculiar velocity. Because there is power on scales below the grid spacing of $20\,h^{-1}$ Mpc, there is not perfect agreement between the two velocity fields. Indeed, these data show a form of velocity bias in the sense that the true velocities are a multiplicative factor higher than the velocities interpolated from the grid. A least-squares fit yields $V_{true} = (1.80\pm0.01)V_{grid}$ with no significant zero-point term, with a scatter about this best-fit line of $274\,$km s$^{-1}$. There is no correlation between the difference between true and interpolated peculiar velocity, and the mass of the cluster. The right-most panel makes this same comparison for our Local Group candidates, and one sees exactly the same trend; least-squares fits to the points shown yield the same slope and scatter, within the errors (there are 16,000 clusters plotted, but only 10,000 Local Group candidates, thus the scatter in the former appears larger). That is, the velocity bias seen here between true and interpolated peculiar velocity is not confined to clusters alone, but is found for more typical $N$-body points.

For the present work, however, we are less interested in the properties of the individual clusters than in their bulk flow. Does the bulk flow measured with clusters agree with that of the field? We placed ourselves on each cluster in the simulation in turn, and calculated the mean peculiar velocity of all clusters within a $150\,h^{-1}$ Mpc sphere around it. In Figure 3, we plot the difference of this quantity and the mean peculiar velocity of the grid points within the same sphere, as a function of the cluster peculiar velocity. Panels a, b, and c are the components of the bulk flows in the three coordinate directions, and panel d is the sum of the squares of the three. The agreement between the the cluster and grid peculiar velocities is excellent, with an rms scatter of only 18 km s$^{-1}$ between them (notice the very different scales on the ordinate and abscissa); any velocity bias between the clusters and the field on these smoothing scales is very small. We find similar results for smaller smoothing scales, although the scatter increases as the mean number of clusters per sphere becomes small. We find similarly good agreement between the bulk flow of the grid points, and that defined by random dark matter particles.

However, this figure shows a disturbing trend: the rms value of the bulk flow itself is a highly anisotropic quantity, being three times larger in the X direction (187 km s$^{-1}$) than in the Y (78 km s$^{-1}$) or Z (60 km s$^{-1}$) directions. Linear theory (Eq. 1) predicts 71 km s$^{-1}$ in each dimension. The anisotropy is due to the finite number of modes in the box on the $150\,h^{-1}$ Mpc sphere; an $800\,h^{-1}$ Mpc box is not a fair sample of the universe, when one is concerned with bulk flows within spheres of $300\,h^{-1}\,\mathrm{Mpc}$ diameter! Indeed, if we subtract from the velocity field the contributions from scales larger than $200\,h^{-1}\,\mathrm{Mpc}$, the bulk flow becomes isotropic (and drops substantially in amplitude). We address this problem in our simulation procedure, which we now detail.

## 3. SIMULATION PROCEDURE

We wish to select many Monte-Carlo realizations of the LP data set from the $N$-body simulations. However, our simulations have a finite volume: a box of $800\,h^{-1}\,\mathrm{Mpc}$ on a side has a volume only $\sim 36$ times larger than that of the LP sample itself. This means that we are subject to the finite number of modes sampled on the largest scales, and thus will tend to underestimate the variance of the bulk flows. We saw this directly in Figure 3; another illustration is shown in Figure 4. In the upper panel, the solid curve shows the rms bulk flow within a sphere of radius $150\,h^{-1}\,\mathrm{Mpc}$ due to that part of the power spectrum on scales larger than $k$:

$$\langle V^2(<k)\rangle^{1/2}_{150\,h^{-1}\,\mathrm{Mpc}} = \frac{H_0\,f(\Omega_0,\Lambda_0)}{2^{1/2}\pi}\left(\int_0^k dk\,P(k)\widetilde{W}^2(k)\right)^{1/2} \qquad (5)$$

(compare with Equation 1). The vertical dashed line is the fundamental mode of the box, with $k = 2\pi/800\,h^{-1}\,\mathrm{Mpc}$. There are substantial contributions to the bulk flow from these large scales. The other five curves correspond to our other models, as indicated in the figure caption. Much of the contribution to the bulk flow on a given scale comes from scales still much larger; large-scale flows arise not from any one structure, but from the sum of fluctuations on all scales. The lower panel similarly shows the cumulative contribution to the density fluctuations within spheres of radius $150\,h^{-1}\,\mathrm{Mpc}$:

$$\left\langle\left(\frac{\delta M}{M}\right)^2(<k)\right\rangle^{1/2}_{150\,h^{-1}\,\mathrm{Mpc}} = \frac{1}{2^{1/2}\pi}\left(\int_0^k dk\,k^2 P(k)\widetilde{W}^2(k)\right)^{1/2} \qquad ; \qquad (6)$$

because of the two extra powers of $k$ in this expression, these curves cut off much more sharply on large scales than do those in the upper panel.

Two conclusions are immediately apparent: (1) the contribution of waves for scales very much larger than the sample scale must be allowed for to accurately estimate the bulk flow on a given scale and (2) the cosmic variance expected for the bulk flow, even on the large scale of the LP sample, is comparable to the bulk flow itself, as the distribution is Maxwellian. We have only a single $N$-body simulation for each model we consider. Thus in order to model these effects properly we need to make random realizations of the large-scale power, both on scales larger than that of the simulation (because these scales are not included at all in the simulation) and on scales comparable to that of the simulation (because these scales are sampled by only a few waves, meaning that they do not properly represent the cosmic variance we must model). We filter the velocity field (although not the density field) of each $N$-body simulation, and subtract the contribution of waves with wavelength between 200 and 800 $h^{-1}$ Mpc. In each realization of the data, we then add back a random realization of the contribution of all waves with wavelength greater than 200 $h^{-1}$ Mpc, given the power spectrum and linear theory.

In detail, then, we follow the following procedure to generate Monte-Carlo realizations of the LP dataset. We work with a given $N$-body simulation in which the contribution to the velocity field from the large-scale waves has been subtracted, as described above. For each realization:

1. We make a random realization of the large-scale power in the velocity field. In particular, for $N_{step} = 100$ values of $k$ logarithmically spaced between $k_{max} = 2\pi/200\,h^{-1}$ Mpc and $k_{min} = 10^{-6}\,h^{-1}$ Mpc$^{-1}$, we choose a random direction for $\mathbf{k}$. The additional contribution to the peculiar velocity at a point $\mathbf{r}$ in the simulation is a sum over these vectors $\mathbf{k}$ given by:

$$\mathbf{V}_{large\ scale} = \frac{H_0 f(\Omega_0, \Lambda_0)}{(2\pi)^3} \sum_{k} \Re \left( i\delta_k e^{i\mathbf{k}\cdot\mathbf{r}} \right) \hat{\mathbf{k}}\ 4\pi k\Delta k \quad , \tag{7}$$

where $\delta_k$ is a complex number with real and imaginary parts Gaussian distributed with variance given by

$$\mathrm{Var} = \frac{\pi^2 P(k)}{k^2 \Delta k} \quad , \tag{8}$$

and $\Delta k = k\,\mathrm{dex}\left[(\log k_{max} - \log k_{min})/(N_{step} - 1)\right]$ is the step size in $k$.

2. We choose an $N$-body point with peculiar velocity smoothed on a $1\,h^{-1}$ Mpc scale between 520 and 720 km s$^{-1}$, and a local density, Gaussian smoothed on a $5\,h^{-1}$ Mpc scale, between 0.8 and 2.0, in order to mimic the "Local Group". This particle is taken to be the observer.

3. The "observed" redshift of each cluster $i$ in the box (assuming periodic boundary conditions) is calculated in the "heliocentric" frame:

$$cz_i = H_0 r_i + \hat{\mathbf{r}}_i \cdot (\mathbf{V}_i - \mathbf{V}_0 + \mathbf{V}_B) + \epsilon \quad , \tag{9}$$

where $\mathbf{V}_i$ is the peculiar velocity of the cluster itself, $\hat{\mathbf{r}}_i$ is the unit vector pointing in the direction of the cluster, $\mathbf{V}_0$ is the peculiar velocity of the Local Group candidate, $\mathbf{V}_B$ is a vector of amplitude 300 km s$^{-1}$ and random direction (fixed for a given realization), mimicking the correction from the Local Group to heliocentric frame, and $\epsilon$ is a Gaussian-distributed error with standard deviation 184 km s$^{-1}$, which reflects the uncertainty in the measurement of the redshift of a cluster (LP). The velocity field used is that of the filtered $N$-body simulation, plus the random field on large scales (see step 1).

4. The subset of cluster points that have heliocentric redshift less than 15,000 km s$^{-1}$ is noted. A coordinate system is set up in which the Local Group peculiar velocity vector is pointed towards Galactic coordinates (277,30), that is, the direction of the CMB dipole. We reject all clusters that fall closer to the Galactic plane than the limits of the LP sample (defined by drawing limiting lines on the sample as a function of longitude). Galactic extinction still causes the cluster sample to be incomplete at above these limits; we model this with the selection function:

$$P(b) = 10^{0.13(1-\csc|b|)} \quad ,$$

(10)

where the coefficient 0.13 was computed from the LP sample itself using a maximum-likelihood technique (Postman *et al.* 1992 find a much higher coefficient (0.32) for the entire Abell catalog, but extinction effects are not as important for the low-redshift LP sample).

5. We sparse-sample the clusters that remain after this selection procedure by roughly a factor of two, to get a mean number of 114 within the sphere. In order to give more weight in the direction of the Great Attractor and the Shapley Supercluster, where the LP sample shows a dramatic overdensity of clusters, we do not sparse-sample those clusters which fall in the area of sky $300 < l < 360°$, $20 < b < 40°$. This does not have a strong effect on the derived bulk flow (the clusters are merely tracers of the velocity field) but this and the careful treatment of the extinction effects in the previous section properly mimic the shape of the error ellipsoid seen in the real data; see below. This is then our mock observational sample. The Local Group redshift of each cluster is noted (Eq. 9 without the $\mathbf{V}_B$ term). Each cluster is assigned a logarithmic slope $\alpha$ of the surface brightness profile of its BCG, drawn randomly from the LP sample. An absolute magnitude in $R_c$ within a metric radius of $10h^{-1}$ kpc is calculated using the observed quadratic relation:

$$M_{R,i} = -20.90 - 4.42\alpha_i + 2.76\alpha_i^2 + \epsilon \quad ,$$

(11)

where $\epsilon$ is a Gaussian-distributed error with standard deviation 0.24 magnitudes. The angular size of the $10\,h^{-1}$ kpc aperture depends on the distance assumed for the cluster. Following LP, we tabulate apparent magnitudes placing clusters at their Local Group redshift distance, requiring an appropriate aperture correction:

$$m_{R,i} = M_{R,i} + 2.5\alpha_i \log_{10}(H_0\,cz_{LG,i}/r_i) + 5\log_{10}(r_i/10\,\text{pc}) \quad .$$

(12)

At this stage, we have a realization of the LP sample: 114 clusters with positions on the sky, Local Group redshifts, $\alpha$, and apparent aperture magnitudes for the BCGs.

6. We now fit for a bulk flow exactly as do LP. Our model is that the peculiar velocity field is described by a bulk flow and small-scale velocity dispersion (cf., the discussion in SCO). We place the clusters at the distance indicated by their Local Group redshifts, and fit the data for a quadratic $L - \alpha$ relation as in Equation 11, via least-squares (i.e., equal weight per cluster). The residuals from the best-fit relation $\delta M$ correspond to a radial peculiar velocity:

$$V_p = cz_{LG} \left( 10^{0.4\delta M/(2-\alpha)} - 1 \right) \quad . \tag{13}$$

A bulk flow is fit to all the data by minimizing $\chi^2$, (i.e., weighting by the inverse square of the errors), where the error for each peculiar velocity is given by $\sigma cz_{LG}/(2 - \alpha)$, where $\sigma$ is the rms value of $\delta M$. This mimics the $n = 2$ solution of LP. With the best-fit bulk flow, the distances to each object are updated, apparent magnitudes recalculated (taking into account the aperture correction, Eq. 12), the $L - \alpha$ relation remeasured, and the bulk flow fitted again. This process is repeated until convergence, typically taking four iterations. This bulk flow is that as measured in the Local Group frame; the motion of the Local Group itself (which we assume is known exactly from the measured dipole of the CMB) is added to this to give the final dipole in the CMB frame. At this stage in their analysis, LP correct their result for two effects: error bias, due to the fact that the derived bulk flow is positive definite, and therefore is biased upwards by errors, and geometric bias, due to a coupling between the dipole moment of the sample geometry and the velocity dipole, because the $L - \alpha$ relation and the dipole are fit simultaneously. We do *not* correct the dipole measured from our simulations for these biases, as we will compare biased results from the LP sample and from the simulations; this is a valid approach to the extent that our simulations duplicate the geometry and error distribution of the real data.

7. Steps 1–6 are repeated 500 times for each $N$-body simulation, and statistics are accumulated on the estimated bulk flow in the CMB frame.

## 4. RESULTS

Before comparing the derived bulk flows directly with the simulations, we examine their characteristics. The upper panel of Figure 5 shows the sky distribution of the LP clusters, the middle panel shows a single realization of the sample drawn from the SCDM model, and the lower panel shows a realization of the New PBI model. Note that the global selection of objects on the sky appears to be about right; in particular, the zone of avoidance in the simulation looks quite similar to that in the real data, and matches the data in an important respect which we will discuss below. However, the real cluster sample is appreciably more clustered than is the SCDM simulation. This is because the SCDM cluster-cluster correlation function is substantially weaker than that of the real universe. The cluster correlation function in the New PBI model is a closer match to the observed correlation function (Cen & Bahcall 1994), and indeed looks more like the real data. It would be interesting to compare the cluster correlation function of the LP dataset and the Monte-Carlo simulations directly as a further test of models (cf., Bahcall & Cen 1992).

In Figure 6, we compare the amplitude of the bulk flow found from each realization of the LP sample with the "right answer", that is, the bulk flow fit to the clusters in the same volume without distance errors. Essentially no correlation is seen: in the case of the SCDM model, the small signal is completely swamped by the photometric distance errors. As we shall see below, the scatter is somewhat broader than that predicted naïvely from propagation of errors, mostly due to the coupling of errors of the bulk flow and the distance indicator relation.

We can now compare the LP results directly to the simulations. We use the same code as for the simulations to fit for a bulk flow directly from their data (as given in Table 3 of their paper); we find a CMB bulk flow of $773 \, \mathrm{km \, s^{-1}}$ towards $l = 355°$, $b = 50°$ (without error bias or geometric bias correction). This differs slightly from the value reported by LP ($806 \, \mathrm{km \, s^{-1}}$ towards $l = 343°$, $b = 52°$ for inverse square weighting before error bias correction) because we strictly limit the sample to heliocentric redshift $< 15{,}000 \, \mathrm{km \, s^{-1}}$ (LP include five clusters at higher redshift associated with the Shapley Supercluster, as Shapley appears at the sample edge), reducing the sample to 114 clusters. In addition, we assume that the surface brightness profiles are strictly power laws (i.e., $\alpha$ is independent of radius for each cluster), while LP redo the aperture photometry from the images themselves on each iteration of their fit (cf., Postman & Lauer 1994). The horizontal dashed line in Figure 6 shows the measured bulk flow; there is an appreciable fraction of points lying above this line.

The upper panel of Figure 7 shows the results of 500 realizations of the LP bulk flow carried out using the standard CDM simulation. The histogram shows the differential distribution (on a linear scale, given on the right-hand side) of calculated bulk flows in the CMB frame, while the smooth, monotonically falling curve is the cumulative distribution (on a logarithmic scale, given on the left-hand side); giving the fraction of realizations which show a bulk flow larger than the value indicated on the ordinate. The vertical dashed line gives the observed bulk flow of $773 \, \mathrm{km \, s^{-1}}$. This value is unusual, but not unheard of in a Standard CDM universe; 11.2% of the realizations find a bulk flow larger than this value. Of course, this is not because the bulk flow on these scales is really so large; rather, it is due to the appreciable scatter in the peculiar velocity estimation, as Figure 6 shows. This becomes clear in Figure 8, which plots the bulk flow velocity estimated from samples of 114 randomly distributed clusters (although still with the selection function of Equation 10 and the Galactic latitude cut), whose peculiar velocities in each component are Gaussian distributed with an rms of $300 \, \mathrm{km \, s^{-1}}$, with no bulk flow component. The tail of large derived bulk flows in this case is as extensive as that seen in the case of SCDM, meaning that for the LP data set, an intrinsic bulk flow would have to be appreciably larger than that predicted in the SCDM model in order that it appear outside the tail due to noise alone. Indeed, one can show in a few steps of algebra that given measured distances for a uniformly distributed full-sky sample of $N$ clusters to radius $R$, with fractional errors in the measured distances $\Delta$, the $1\sigma$ uncertainty in the derived amplitude of the bulk flow, with inverse square weighting, is $\sqrt{1/N}\,\Delta R$. For the present sample ($N = 114$, $\Delta = 0.17$, and $R = 15{,}000$ km s$^{-1}$), this gives $240$ km s$^{-1}$. This implies that the error-free bulk flow of a model would have to be greater than $\sim 480$ km s$^{-1}$ in the CMB frame in order to be detected with confidence ($2\sigma$) in the LP experiment.

However, as LP emphasize, the current sample is not isotropic; in particular, because of the zone of avoidance, there is higher sensitivity to detecting the component of the bulk flow towards the Galactic poles than the orthogonal components. This is quantified by the covariance matrix of the bulk flow calculation, whose inverse is given by:

$$M_{lm} = \sum_{\text{clusters } i} \frac{\hat{r}_{i,l}\hat{r}_{i,m}}{(\Delta r_i)^2} \quad , \tag{14}$$

where the sum is over the clusters in the sample, $\hat{r}_{i,l}$ is the $l$th component of the unit vector towards the $i$th cluster, and $\Delta r_i$ is the error in the measured distance to the galaxy. The fractional error of the distance $\Delta$ is given by $\sigma/(2-\alpha)$, where $\sigma$ is the scatter measured in the $L-\alpha$ relation from each realization. For the LP sample itself, the eigenvalues of the inverse of $M$ (i.e., the length of the axes of the error ellipsoid) are in ratio $1 : 1.3 : 1.9$, and in fact, the observed bulk flow is directed almost exactly along the *short* axis of the ellipsoid. That is, the statistical significance of the observed bulk flow is much larger than a naïve estimate based on its total amplitude and the isotropic error would indicate. We quantify this by defining a $\chi^2$ statistic of the bulk flow relative to the null hypothesis of no bulk flow at all:

$$\chi^2 = V_l V_m M_{lm} \quad , \tag{15}$$

where $V_l$ is the $l$th component of the derived bulk flow, and Einstein summation is assumed[3]. For the LP sample, $\chi^2 = 14.14$, corresponding roughly to a $3\sigma$ detection given 3 degrees of freedom (the three components of the bulk flow) (Press *et al.* 1992). Again, this result differs from the value of $\chi^2 = 20.6$ found by LP, because the latter authors find a slightly different bulk flow vector, correct for geometric bias, and derive the covariance matrix by Monte-Carlo methods rather than by Equation 14. For each of the Monte-Carlo simulations, we can calculate the quantity $\chi^2$ for the bulk flow found, again using Equation 14 to determine the covariance matrix in each case. The lower panel of Figure 7 shows the results for the Standard CDM simulation. Again, the histogram shows the differential distribution of $\chi^2$, and the monotonic smooth curve shows the cumulative distribution on a logarithmic scale. The dotted curve shows the expected cumulative distribution of $\chi^2$ for three degrees of freedom given the null hypothesis (no bulk flows). The true distribution is more extended, due to the real bulk flows in the models and the non-Gaussian nature of the error distribution (cf., Figure 8; even with no intrinsic bulk flows, the observed distribution is more extended than the expected $\chi^2$ distribution). The vertical dashed line is drawn at the observed value of $\chi^2 = 14.14$; only 13 realizations out of the 500 (2.6%) show a value of $\chi^2$ larger than that observed. Thus the LP data rule out SCDM at the $97.4 \pm 0.7\%$ confidence level, the error bar given by Poisson statistics. In practice, successive sets of 500 simulations for a given model gave confidence levels that differ at the 2% level, which is perhaps a more realistic estimate of the uncertainty in the confidence level.

---

[3] Equations 14 and 15 give a proper $\chi^2$ statistic to the extent that the measurement errors are Gaussian. Postman & Lauer (1994) show that the scatter around the $L-\alpha$ relation is accurately Gaussian, implying a log-normal distribution for the distance errors. However, whatever biases this may cause will be the same for the observations and the simulations. Moreover, the Central Limit Theorem will make the errors in the components of the bulk flow more accurately Gaussian than the errors in the individual distances.

The $\chi^2$ statistic differs from the amplitude of the bulk flow statistic in the upper panel only to the extent that the error ellipsoid is anisotropic (that is, if it were isotropic, the two statistics would be redundant). Thus it is vitally important that $a$), the errors are modeled correctly in the simulations, and $b$), that the error ellipsoid have the same shape in the simulations as in the real data. To check the first point, we have examined the distributions of the derived scatter of the $L-\alpha$ relation in the SCDM realizations, and found that it indeed has a mean of 0.24 magnitudes (the input value), with a Gaussian distribution with a scatter of 0.017. Moreover, the rms weighted distance of clusters in the simulations is $8320\,\mathrm{km\ s^{-1}}$, close to the observed value of $8665\,\mathrm{km\ s^{-1}}$. As described above, the simulations include the effect of an excluded zone and Galactic extinction, and are sampled more densely in the direction of the Shapley Supercluster. This causes an anisotropy in the error ellipsoid as in the real data: the distribution of ratios of the largest to smallest axes of the error ellipsoid in the SCDM simulations have a mean of 1.83 and a standard deviation of 0.19, close to the value of 1.91 seen for the LP sample itself. No correlation is seen between $\chi^2$ and this ratio. Indeed, our results are insensitive to exactly how the excluded zones are treated; we experimented with a variety of schemes to match the spatial distribution of the LP sample, and consistently found SCDM to be ruled out at the 95% confidence level or better.

LP correct their calculated bulk flow for error bias (due to the fact that the amplitude of the bulk flow is a positive definite quantity) and for geometric bias due to the fact that the sample is not isotropic. We have corrected neither the observations nor the simulations for these biases, preferring to make direct comparisons of uncorrected quantities. This is a valid procedure to the extent that the biases (which depend on the distribution of clusters and the scatter in the distance indicator relation) are the same for both models and data. In any case, the error bias is *not* subtracted by LP for the $\chi^2$ statistic, and their geometric biases are small. Moreover, calculations of the geometric biases from the simulations are in good agreement with those listed in LP.

Comparison of Figures 7 and 8 indicates that any model which exhibits intrinsic bulk flows on the scales probed by the LP sample which are small relative to the noise will be ruled out at a similar confidence level by the LP observation. Our results for all models are shown in Table 2, listing the fraction of realizations in each model which gave bulk flows and $\chi^2$ values larger than those found in the real data. The second column of the table lists the actual bulk flow within the spheres (i.e., without any errors applied) found by fitting to the radial components, and averaging over realizations. As we discuss in the following section, this gives different results from summing up all three components of the peculiar velocity in the presence of flows on intermediate scales. The linear theory prediction via Equation 1 for a sphere of radius $150\,h^{-1}$ Mpc (cf., Figure 4) is shown in the next column. Next follows the mean bulk flow over realizations, now with the errors applied. If the latter numbers were corrected for error bias following LP, we would find results in rough agreement with the error-free bulk flows, although with huge scatter.

The fraction of realizations giving bulk flows and $\chi^2$ values larger than those observed follow in the next two columns. All models are ruled out at better than the 94% confidence level by the $\chi^2$ statistic, with HDM faring the best, and LCDM the worst (98%). Note that all models fare roughly equally as well; the differences in true bulk flows between models is lost in the noise. Even the New PBI model, with its much larger bulk flows on large scales (cf., Figure 4) does no better than the others.

LP carry out a number of tests of the robustness of their result. Among other things, they show that the derived bulk flow is robust to various cuts in the sample. We carried out one such experiment in the SCDM simulation: we cut the sample at $cz = 10,000$ km s$^{-1}$, to increase the mean signal-to-noise ratio of each individual peculiar velocity (leaving 43 clusters; in this case, the rms scatter in the $L - \alpha$ relation is 0.21 mag). In this case, $\chi^2$ stays high at 13.81, but the much smaller number of clusters increases the noise of the estimate, and 6.6% of the realizations are able to match the observed value.

Finally, to see how sensitive the results were to the exact geometry of the LP sample, and the positioning of observed points on clusters found in the simulations, we did a series of realizations, placing our "observations" at positions corresponding to the actual positions of the LP clusters, within the SCDM simulation. Thus unlike the simulations above, the data points do not correspond to actual clusters in the simulations, but they exactly reproduce the LP geometry and therefore the LP error ellipsoid. In practice, the peculiar velocity corresponding to each point was interpolated from a grid of spacing $20\,h^{-1}$ Mpc (see §3) from which components of the velocity field on a scale larger than $200\,h^{-1}$ Mpc had been subtracted; the large-scale waves were then added back in using Equation 7. Measurement errors were added and bulk flows measured exactly as in the simulations above. The results are given in the last line of Table 2; the constraint on the model from the $\chi^2$ statistic is comparable to what we found with the cluster selection.

## 5. DISCUSSION AND CONCLUSIONS

The velocity field model which we have assumed in fitting bulk flows to the data and the simulations consists of a constant large-scale flow measured with a distance indicator with appreciable scatter; components of the velocity field on intermediate scales are not explicitly taken into account. In fact, the velocity field in general has components on all scales, which means that the bulk flow derived from a data set may not be equal to our intuitive notion of what it should be. Imagine an observer in a homogeneous universe, and add a single cluster displaced from the origin, which gravitationally attracts the particles around it. The mean bulk flow of a sphere centered on the observer containing the cluster is non-zero, despite the fact that there is no source of gravity from outside the sphere (Juszkiewicz, Vittorio, & Wyse 1990); indeed, this component of the bulk flow is proportional to the distance of the center of mass to the origin. One can show in linear theory that the bulk flow due to the fact that the center of mass of a sphere does not coincide with its geometric center is given by:

$$\left\langle V^2 \right\rangle^{1/2} = \frac{H_0\, f(\Omega_0, \Lambda_0)}{2^{1/2}\, \pi} \left( \int dk\, P(k) j_2^2(kR) \right)^{1/2} \quad . \tag{16}$$

This is plotted against $R$ in Figure 9, together with the total expected rms bulk flow; this effect becomes increasingly important on larger scales. Of course, to the extent that the simulations are realistic, this effect is included to the same extent as in the real data.

A further related effect is that the bulk flow found by fitting our model to the radial components of the peculiar velocities is quite different from that calculated by summing up the peculiar velocity vectors (if we could somehow observe the transverse components). One could get around the latter problem if one had *a priori* knowledge of the velocity correlations on different scales (in the sense of Górski (1988), for example), and included this in the least-squares solution for the bulk flow using radial peculiar velocities alone. For the LP sample, we have seen that random errors due to the finite scatter of the distance indicator relation dominate small flows completely, and the additional refinement of modeling the velocity correlations is unlikely to make a difference. This however is an avenue we will pursue with peculiar velocity data sets at lower redshifts, with higher signal-to-noise ratio per point.

In SCO, we compared the bulk flow found in various peculiar velocity data samples with the small-scale velocity dispersion; the ratio of the two, the Cosmic Mach Number, gives a measure of the relative amounts of large-scale and small-scale power. The LP sample does not allow a straightforward measure of the small-scale velocity dispersion, as the distance indicator relation is calibrated directly from the data set itself, meaning that the intrinsic scatter in the relation is added in quadrature with the effects of peculiar velocities. However, the errors in peculiar velocities due to the former scale proportional to the distance of each cluster, while the latter are nominally independent of distance. LP used this fact to estimate that the small-scale velocity dispersion was less than $250 \, \text{km s}^{-1}$ (1-d). We have calculated the corresponding limits in our Monte-Carlo simulations; unfortunately, the small-scale velocity dispersion is much too noisy a statistic to make this a powerful test of models.

LP concentrated on measuring the bulk flow, or dipole moment, from their sample; indeed, their sample was designed explicitly for this measurement. There is no reason in principle not to measure other multipoles from the data and simulations. Because the distance indicator is calibrated directly from the data, it would appear that the monopole is identically zero. However, the distance indicator relation is determined giving each cluster equal weight, while the multipole solution gives weights inversely proportional to the square of the estimated errors. Thus there is a residual monopole in the data, but it is very strongly coupled to the geometry of the sample (in particular, its radial profile). Similarly, the quadrupole moment of the data can be measured, but it couples strongly to the quadrupole moment of the cluster distribution (cf., Figure 5); that is, the geometric correction for this statistic is very large. This strong coupling means also that there is strong covariance between the parameters of the $L - \alpha$ relation itself and these moments, meaning that they have large variances. Indeed, simulations showed that fitting for the monopole, dipole, and quadrupole simultaneously gave a much weaker constraint on the theories than did the dipole alone.

Two other papers have discussed the LP dataset in a theoretical context. Feldman & Watkins (1994a,b) have carried out an analysis similar to that here, although they put more emphasis on an analytic approach, and do not take into account the anisotropy of the error ellipsoid, which led us to the statistic with the most power to rule out cosmological models. Their conclusions are broadly consistent with ours, that the amplitude of the LP bulk flow (not the $\chi^2$ statistic!) rules out models which match observations of galaxy clustering, and the COBE anisotropies, at the 95% confidence level. Tegmark, Bunn, & Hu (1994) extended the formalism of Juszkiewicz, Górski, & Silk (1987) to show that the LP result is inconsistent at the 95% confidence level with observations of small-scale CMB anisotropies *independent* of power spectrum.

None of the models we have examined are favored by the LP dataset. Some of them are ruled out on other grounds: HDM models have been out of favor for some time due to their late epoch of galaxy formation and the overly massive clusters (Cen & Ostriker 1993a and references therein). The Tilted CDM model forms small-scale structure late, and fails to explain the absence of Gunn-Peterson absorption in high-redshift QSO spectra (Cen & Ostriker 1993b); it is also disfavored by bulk flow data on smaller scales (Mucciacia *et al.* 1993). The most severe problems that the SCDM model faces are on small scales: it predicts a velocity dispersion on small scales appreciably higher than that observed, even allowing for a velocity bias of 0.8 for galaxies (e.g., Cen & Ostriker 1992); it also over-predicts the power spectrum on small scales (Fisher *et al.* 1993). The PBI model predicts bulk flows on $150 \, h^{-1} \, \text{Mpc}$ scales 50% higher than those of the other models, but this is still washed out by the noise; the LP observations rule it out to roughly the same confidence level as the other models. The cluster correlation and mass functions are also a sensitive discriminant of models (cf., Bahcall & Cen 1992); the New PBI model is the one that best fits the data.

The Monte-Carlo simulations carried out in this paper will give misleading results if the assumed scatter in the $L - \alpha$ relation is underestimated. The higher the true error, the greater the tail of high inferred bulk flows. However, the $\chi^2$ distribution is not broadened considerably by increasing the input scatter, as the covariance matrix in Equation 14 is normalized relative to the inferred scatter. Moreover, we saw that the scatter measured from the simulations accurately reproduced the input value, meaning that it is unlikely that the true underlying dispersion is much larger than that quoted by LP. Indeed, a Monte-Carlo experiment with the input scatter set at 0.30 mag showed no realizations out of 500 with a measured scatter as small as that observed, and even with an input scatter of 0.27 mag, only 5% of the realizations showed a scatter as small as that observed.

Given that the LP data set appears to rule out so many models "in one fell swoop" (albeit at the $2 - 3\sigma$ level), one might be tempted to explain their bulk flow away to some subtle systematic effect. It should be pointed out, however, that the LP sample is selected using well-defined criteria, covers the entire sky, has uniform quality data and is self-calibrating. No other peculiar velocity survey can boast all of these qualities. LP have carried out an extensive series of tests of the robustness of their results, ruling out a large variety of possible systematic effects. Moreover, the observations are being extended in two ways. First, TRL and MP have found that velocity dispersion is a second parameter in the $L - \alpha$ relationship, and that including velocity dispersion decreases the scatter of the distance indicator relation from 0.24 to $\sim 0.18$ mag. We have carried out Monte-Carlo simulations of the LP sample using this smaller scatter; if their bulk flow result holds up, then all of the models discussed here will be ruled out at better than the 1% confidence level. This is not surprising; it is the errors which drive the tail of the distributions seen in Figure 7, and the realizations with the largest "measured" bulk flows do not have unusually large "true" bulk flows. Decreasing the distance indicator errors will greatly decrease the extent of the tail. Second, three of us (TRL, MP, and MAS) have started a program to obtain distances to the brightest cluster members of a complete sample of Abell clusters to 24,000 km s$^{-1}$, greatly increasing the effective volume of the survey. This incidentally will allow an independent measure of the scatter in the $L - \alpha$ relation. In addition, various other groups are testing the LP results by measuring independent distances to clusters: Tully-Fisher for spirals (Willick et al. in preparation), $D_n$–$\sigma$ for ellipticals (Colless et al. 1993), and luminosity function fitting to the whole cluster galaxy population (Moore et al. in preparation).

In conclusion, we have done detailed Monte-Carlo simulations of the LP cluster data set, mimicking all relevant aspects of their observations and reduction procedures. We use $N$-body simulations of six cosmological models, spanning the range of currently popular models which assume Gaussian initial conditions. We find that when derived bulk flows from the simulations are properly normalized by the (anisotropic) error ellipsoid, the observations rule out all models at the 95% confidence level and better, and will do so for any model that predicts bulk flows on $150\, h^{-1}$ Mpc scales smaller than $\sim 400$ km s$^{-1}$. Thus the LP data are consistent with the detection of a bulk flow of unprecedented amplitude.

We thank David Weinberg for useful discussions. MAS is supported at the IAS under NSF grant PHY92-45317, and grants from the W.M. Keck Foundation and the Ambrose Monell Foundation. RYC and JPO are supported by NASA grant NAGW-2448, and NSF grants AST90-06958 and HPCC ASC93-18185. MP is supported by NASA grant NAGW-2166.

TABLE 1

THE PARAMETERS OF THE POWER SPECTRA

| Model | $\Omega_0$ [a] | $\sigma_8$ [b] | $H_0$ [c] | $n$ [d] or $m$ [e] | $x$ [f] |
|---|---|---|---|---|---|
| Standard CDM | 1.0 | 1.05 | 50 | 1.0 | 0.0 |
| HDM | 1.0 | 0.86 | 50 | 1.0 | 0.0 |
| PBI | 0.2 | 0.90 | 80 | $-0.5$ | 0.1 |
| $\Omega_0 = 0.3$ CDM | 0.3 | 0.67 | 67 | 1.0 | 0.0 |
| Tilted CDM | 1.0 | 0.50 | 50 | 0.7 | 0.0 |
| New PBI | 0.3 | 1.02 | 50 | $-1.0$ | 0.1 |

[a] All models are spatially flat so that $\Lambda_0 = 1 - \Omega_0$.
[b] rms fraction density fluctuations within $8\,h^{-1}$ Mpc spheres.
[c] Hubble Constant in units of km s$^{-1}$ Mpc$^{-1}$.
[d] Adiabatic primordial spectral index.
[e] Isocurvature primordial spectral index.
[f] Ionization fraction.

TABLE 2

BULK FLOWS IN VARIOUS MODELS

| Model | True Bulk Flow ( km s$^{-1}$ ) | Eq. 1 ( km s$^{-1}$ ) | Measured Bulk Flow ( km s$^{-1}$ ) | % exceeding Bulk Flow | % exceeding $\chi^2$ |
|---|---|---|---|---|---|
| Standard CDM | $126 \pm 67$ | 123 | $486 \pm 244$ | 11.2 % | 2.6 % |
| HDM | $132 \pm 58$ | 147 | $539 \pm 257$ | 17.8 % | 5.0 % |
| PBI | $101 \pm 45$ | 137 | $445 \pm 208$ | 6.2 % | 2.2 % |
| $\Omega_0 = 0.3$ CDM | $72 \pm 31$ | 126 | $461 \pm 209$ | 9.4 % | 1.6 % |
| Tilted CDM | $98 \pm 41$ | 92 | $482 \pm 232$ | 11.0 % | 4.2 % |
| New PBI | $163 \pm 63$ | 184 | $472 \pm 217$ | 10.0 % | 3.4 % |
| Poisson | — | — | $519 \pm 228$ | 13.0 % | 1.0 % |
| Standard CDM[a] | $220 \pm 180$ | 176 | $548 \pm 272$ | 18.6 % | 6.6 % |
| Standard CDM[b] | $181 \pm 86$ | 123 | $441 \pm 211$ | 5.8 % | 2.4 % |

[a] Cut at $cz = 10,000$ km s$^{-1}$.
[b] Sampled with the exact LP geometry.

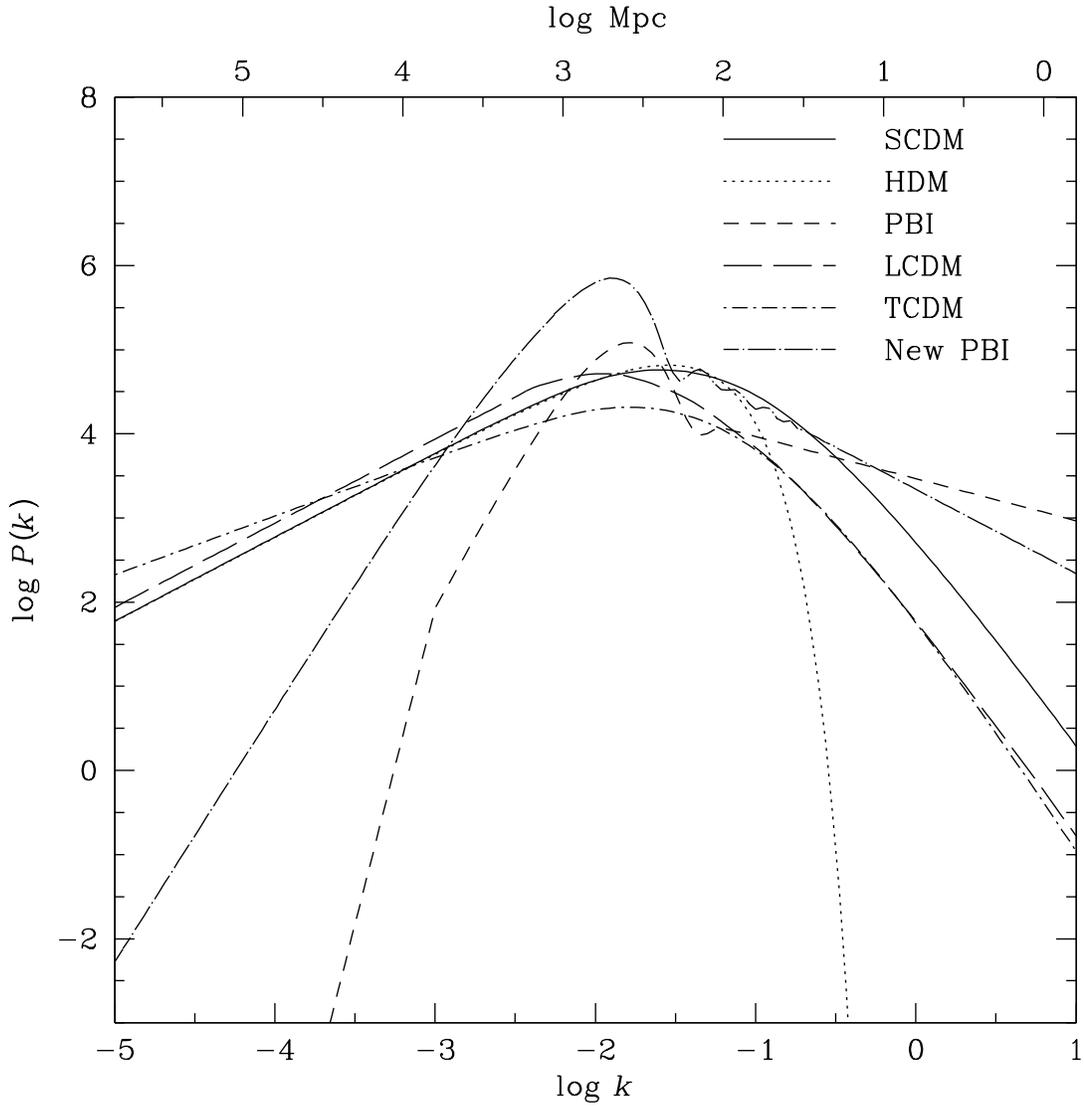

Figure 1. The six power spectra used in this paper. The correspondence between line types and spectrum is given in the upper right corner. All models, with the exception of HDM, have been normalized to the COBE fluctuations at 10°. The upper scale gives the wavelength corresponding to wavenumber $k$, in Mpc.

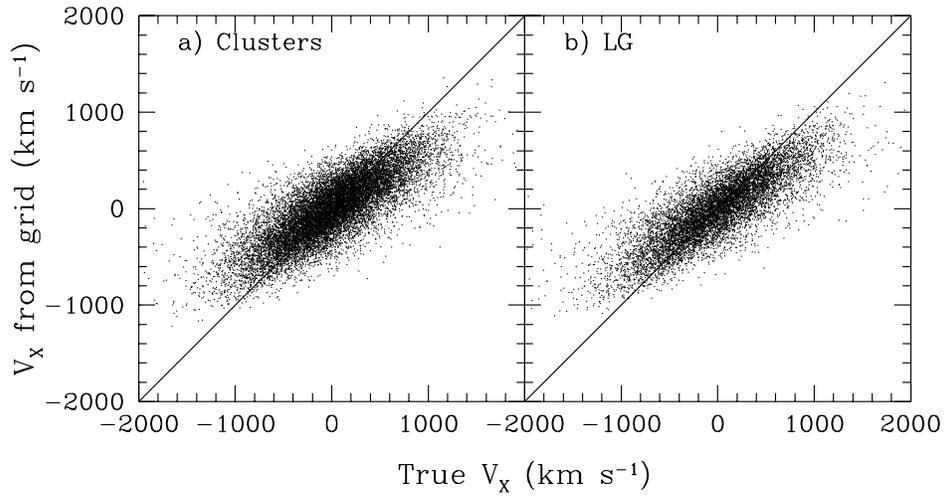

Figure 2a. Correlation between the $x$ component of the peculiar velocity of each cluster in the SCDM simulation (abscissa) and the value of the peculiar velocity interpolated from a Cartesian grid with $20\,h^{-1}\,\mathrm{Mpc}$ spacing (ordinate). Note the systematic deviation from equality (diagonal line). Figure 2b. As in a, for the Local Group candidates. There is no difference in behavior between clusters and randomly chosen $N$-body points.

Bulk Flows in 150 $h^{-1}$ Mpc Spheres

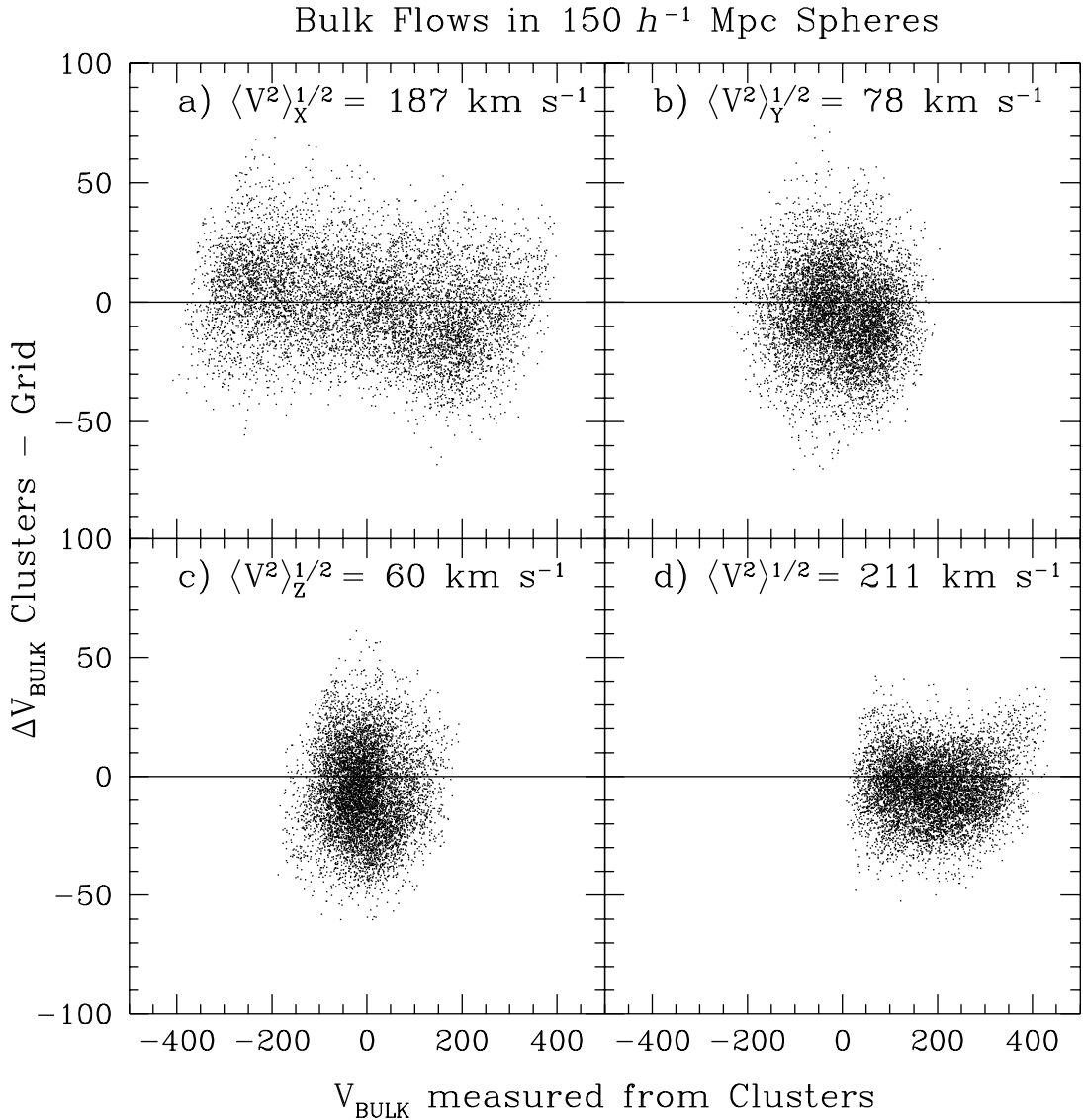

Figure 3. The relation between the bulk flow in 150 $h^{-1}$ Mpc spheres in the SCDM simulation as measured by clusters (abscissa) and as measured from a uniform grid. The agreement between the two is excellent. The panels are the components in the a) X, b) Y, and c) Z directions; the amplitude of the bulk flow is in panel d). Note however the strong anisotropy; the X component shows a dispersion three times larger than the other two components, due to the small number of waves in the box on these scales.

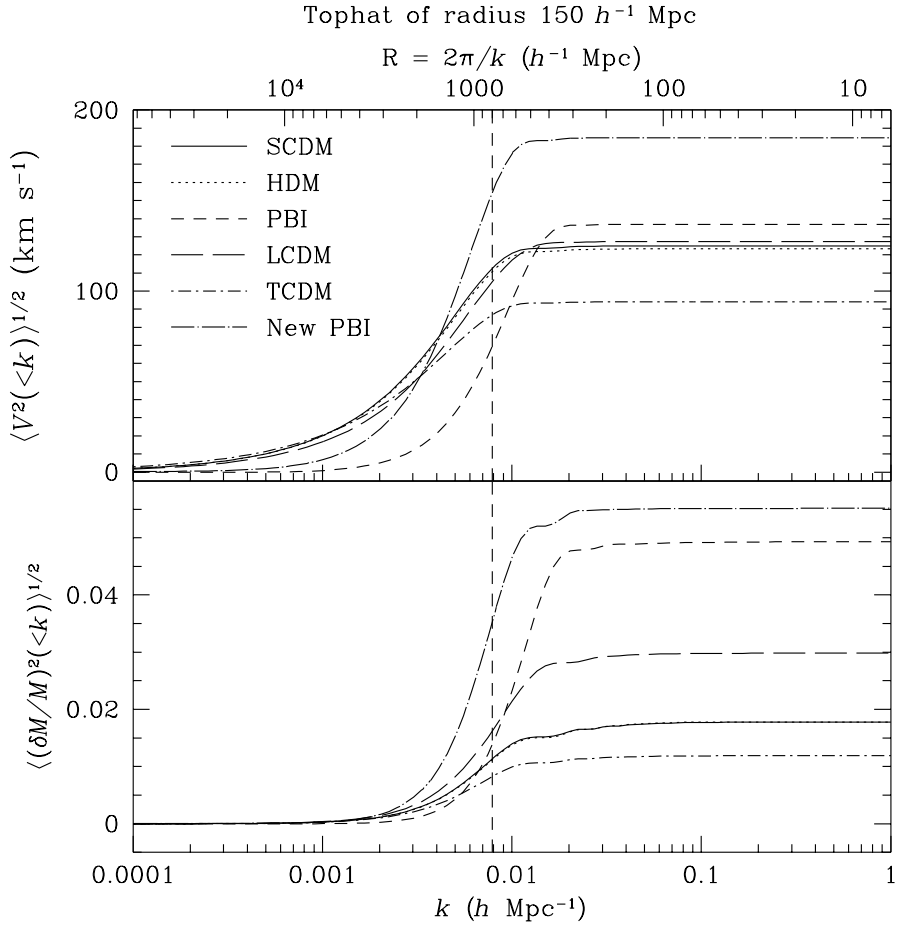

Figure 4. The upper panel plots the cumulative contribution to the bulk flow in a $150\,h^{-1}$ Mpc radius sphere as a function of the wavenumber $k$ assuming linear theory (Eq. 5), for the six models examined. The lower panel shows the cumulative contribution to the mass fluctuations in the same sphere. Note that the latter is contributed by a much narrower range of scales.

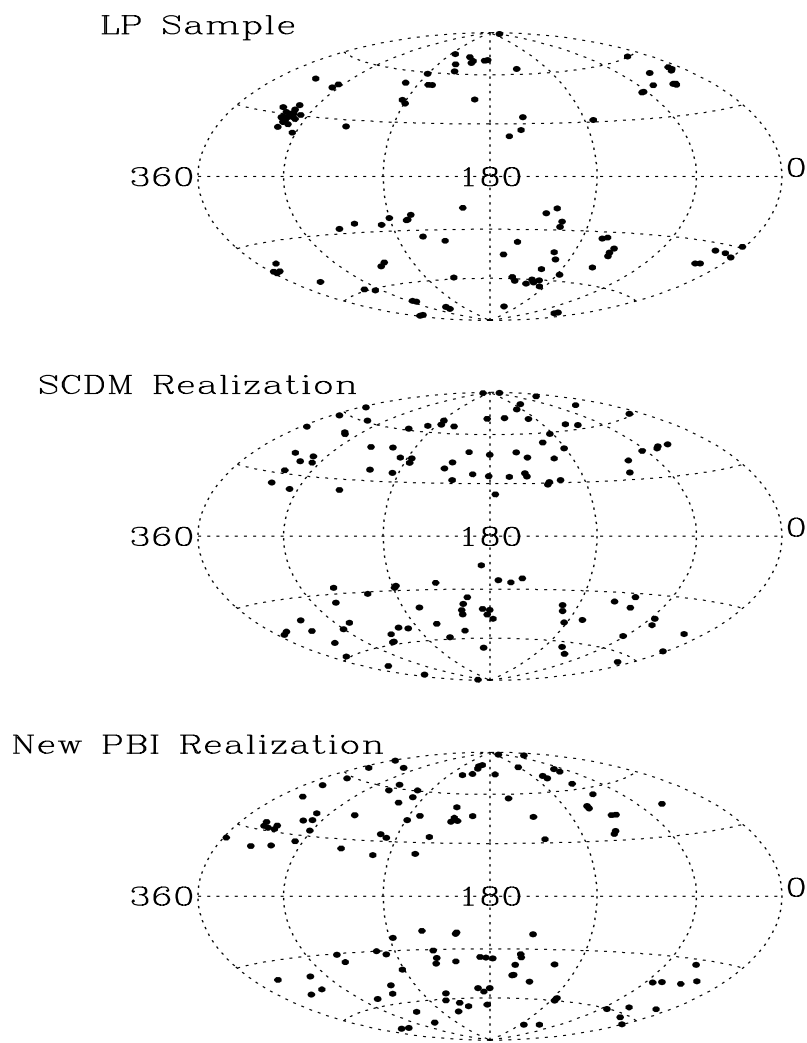

Figure 5. The sky distribution of the Warpfire sample (upper panel), one realization of the sample drawn from the Standard CDM model (middle panel), and the New PBI model (lower panel). Note the greater clustering apparent in the real data, than in the SCDM model; New PBI fares somewhat better.

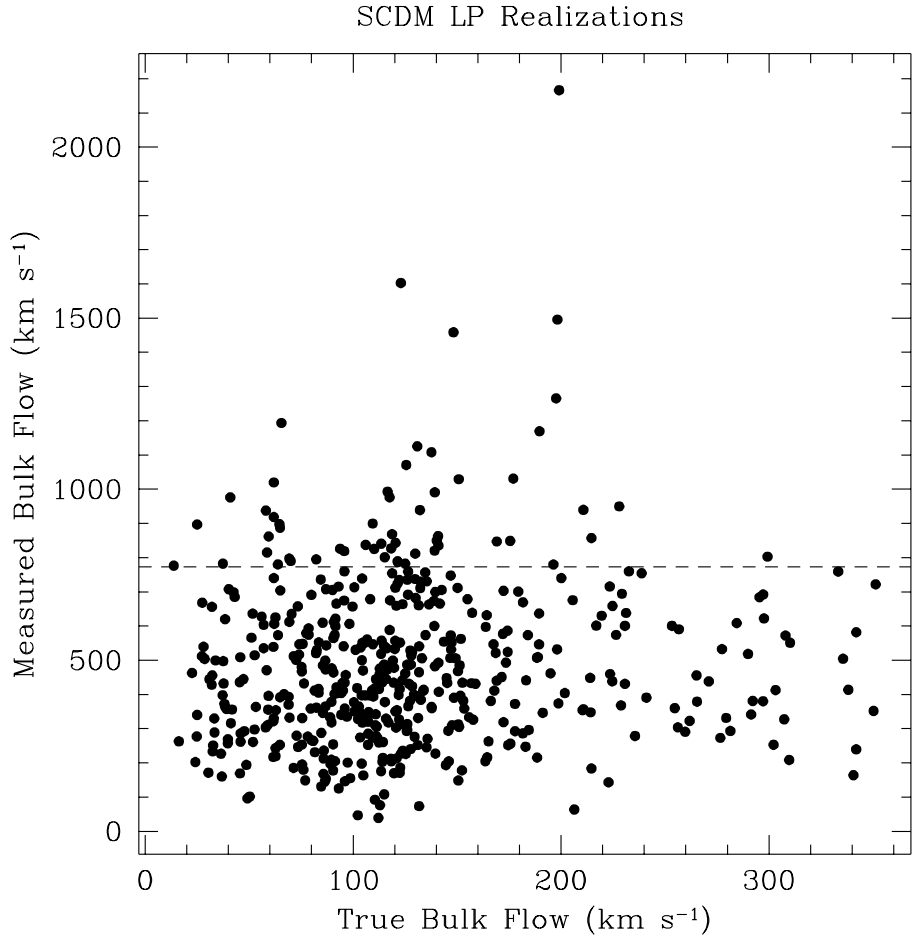

Figure 6. The true bulk flow of clusters in each of the SCDM Monte-Carlo realizations, plotted against the bulk flow inferred from the the simulated noisy data. The true bulk flow is negligible relative to the noise, and no correlation between the two is seen. The dashed line is the LP measurement of bulk flow ($773\,\mathrm{km\,s^{-1}}$), without error bias correction.

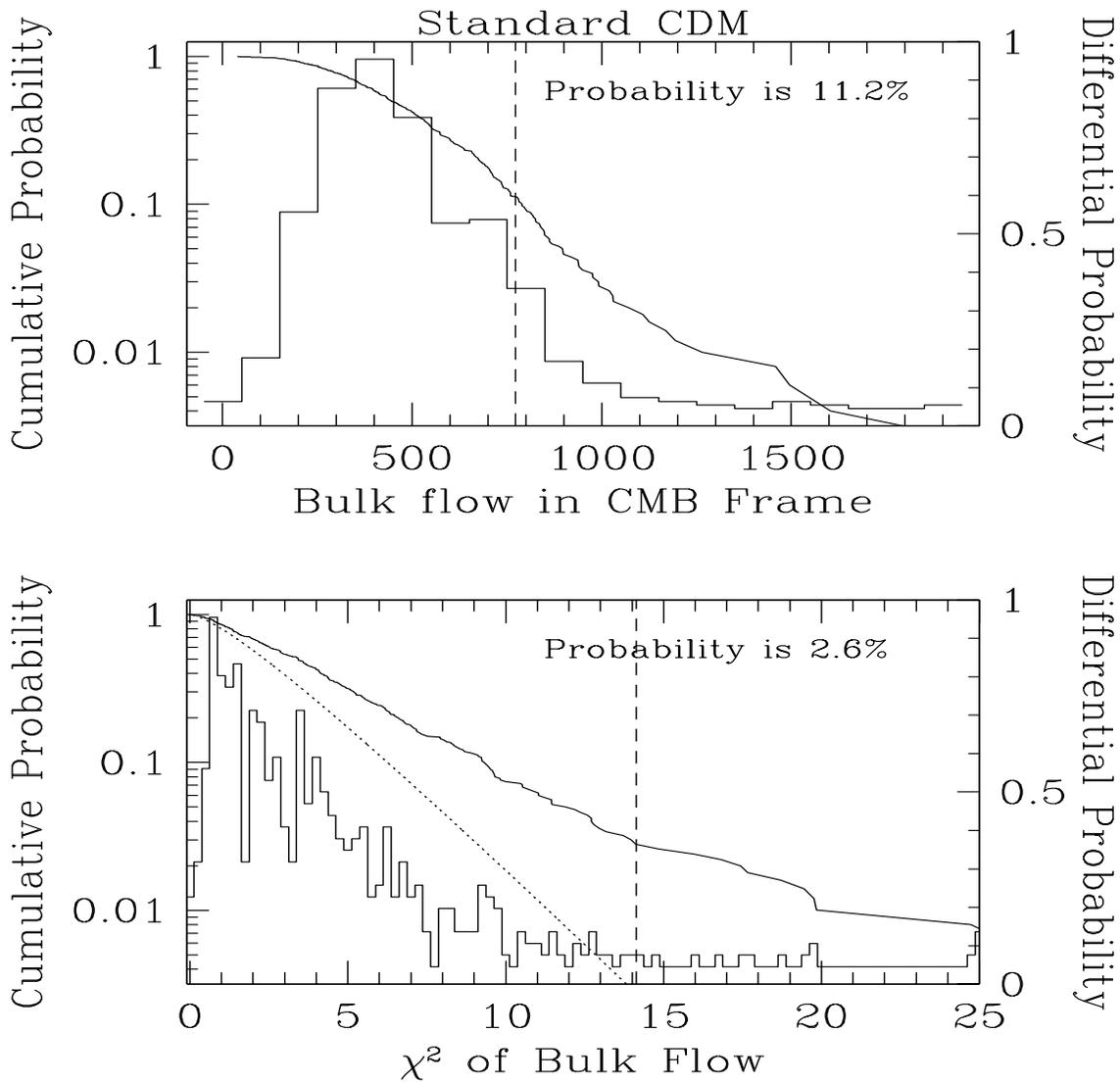

Figure 7. The upper panel gives the differential distribution (histogram) and cumulative distribution (monotonic curve, on a logarithmic scale) of derived bulk flows from realizations of the Warpfire sample from a Standard CDM simulation. The vertical dashed line is the observed bulk flow. The lower panel shows the distributions for the quantity $\chi^2$; the vertical dashed line is the observed value. The dotted line in the expected cumulative distribution of $\chi^2$ with 3 degrees of freedom.

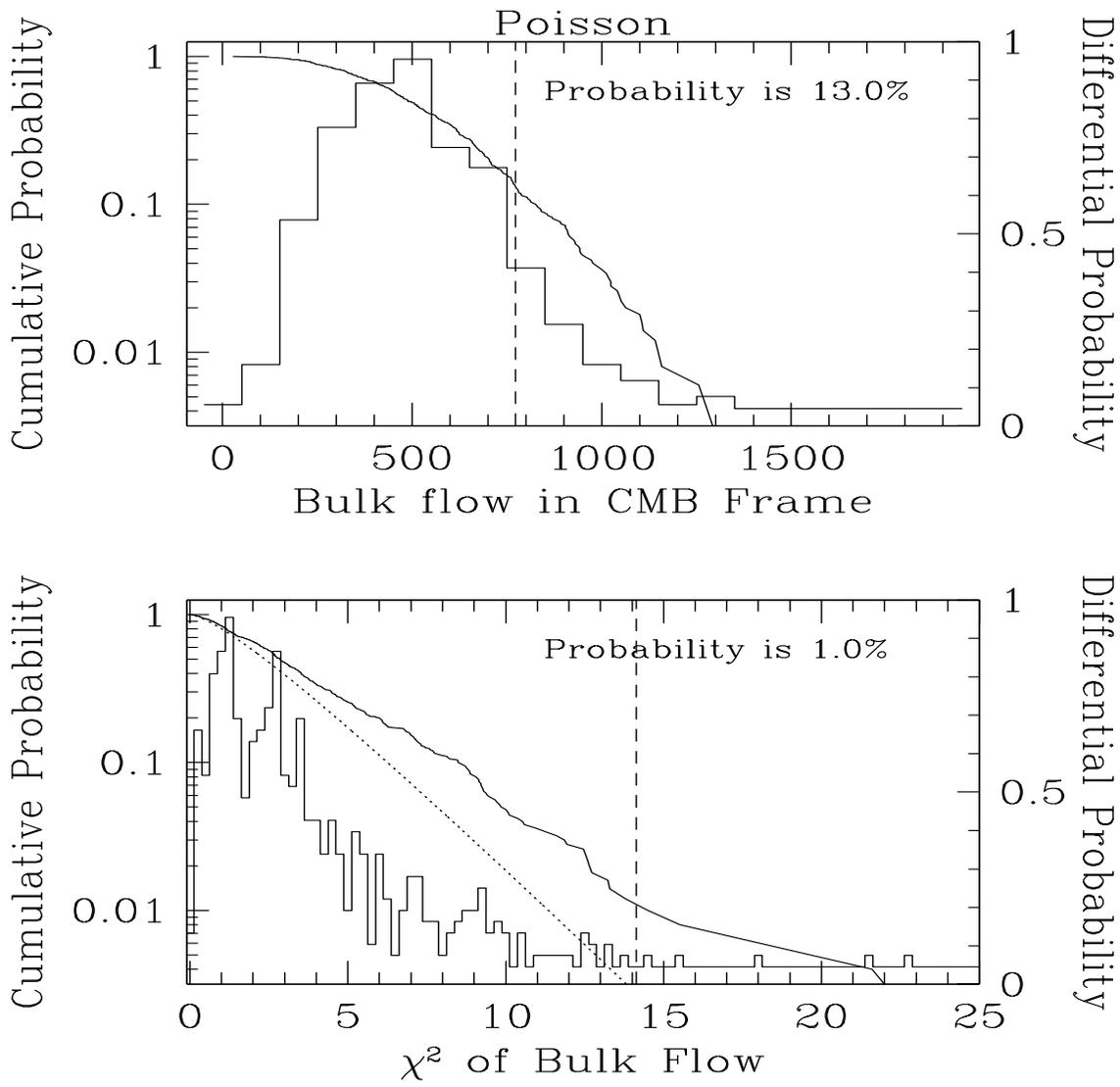

Figure 8. As in Figure 2, for a Poisson distribution.

# Bulk Flows in Spheres; Equations 1 and 16

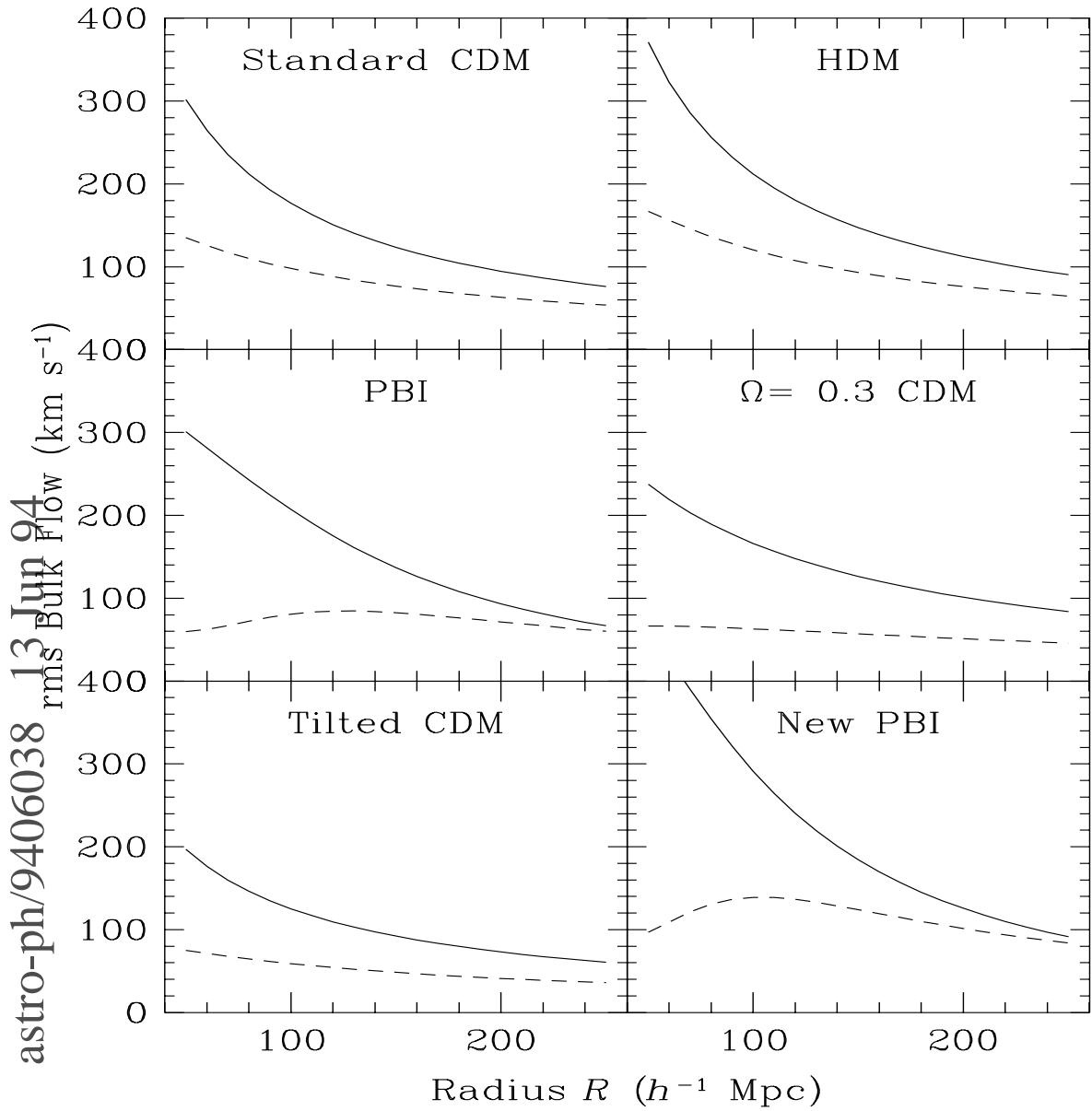

Figure 9. The rms expected bulk flow in spheres of radius $R$ (Equation 1) for six models is shown as solid curves. The component of the bulk flow due to displacement of the center of mass of a sphere relative to its geometric center (Equation 16) is shown as dashed curves.